# Spherical dust acoustic solitary waves with two temperature ions

**Esmaeil ESLAMI[1], and Rasoul BARAZ**,

Department of Physics, Iran University of Science and Technology (IUST), Narmak, Tehran, 16846-13114, IRAN

---

[1]Author to whom all correspondence should be addressed. Email: eeslami@iust.ac.ir

# Abstract

The nonlinear dust acoustic waves in unmagnetized dusty plasma which consists of two temperature Boltzmann distributed ions and Boltzmann distributed electrons in spherical dimension investigated and obtained spherical Kadomtsev–Petviashvili (SKP) equation and shown that the dust acoustic solitary wave can exist in the SKP equation.

## 1. Introduction

Dusty plasma is different than common plasma and consists of heavy dusts as well as ions and electrons and has been observed in planetary rings, asteroid zones, commentary tails, magnetospheres as well as the lower part of the Earth's atmosphere[1-5]. These heavy dusts or grains can cause to create modes such as dust acoustic waves, dust ion acoustic double layer etc. These modes except on damping factor[6] are in low frequency oscillations[7, 8] in the range of a few to tens of Hz.

They were Rao and Shukla[8] who first theoretically predict dust acoustic wave mode in un magnetized dusty plasma and later the experimental observations[9-13] have confirmed the existence of linear and nonlinear feature of both the dust acoustic and dust ion acoustic waves.

Usually, the solitary waves in non-dimensional form or bounded in one dimension is describe by the reduction perturbation technique to give rise to Korteweg–de Vries (KdV) or Kadomtsev - Petviashvili (KP) equation[14-19]. However this scheme is not a realistic situation in space and laboratory experiments and the geometry is better to extended to two or three dimensions[15, 20-22]. In addition, the well-known transverse perturbation introduces anisotropy into system and can be modified the structure of solitary waves[23, 24]. This effect always appears in the higher-dimensional system.

Mamun and Shukla[25] was considered the nonlinear propagation of cylindrical and spherical dust acoustic waves in an unmagnetized dusty plasma by employing the reductive perturbation technique to derive a modified Korteweg–de Vries (KdV) equation. Xue[26] investigated a spherical Kadomtsev-Petviashvili (KP)equation in dusty plasma which consisted of Boltzmann distributed electrons and ions. Then, Mirza et al[27] were studied the dust acoustic solitary waves in the presence of hot adiabatic dust in cylindrical and spherical geometries in an unmagnetized dusty plasma to derive the modified Korteweg–de Vries (mKdV) by using the reductive perturbation technique. Time-dependent cylindrical and spherical dust-acoustic solitary and shock waves propagating in a strongly coupled dusty plasma in the presence of polarization were investigated by Ashrafi et al[28]. The plasma consisting of cold dust grains, super thermal ions and electrons in spherical geometry were investigated by some authors[29, 30]. Nonextensive dust acoustic solitary and shock waves in nonplanar geometry was studied by Sahu and Tribeche[31].

They have concluded that the polarity switch may be attributed to the ion nonextensivity which may originate from the correlation or long-range interactions in the dusty plasma under concern.

In our previous work, we have investigated the effects of two-temperature ions and external magnetic field on the nonlinear propagation of dust-acoustic waves in magnetized dusty plasmas for a non dimensional case. Here we tried to study a more realistic dusty plasma with radial symmetry of an unmagnetized dusty plasma whose constituents are negatively charged inertial dust particles, super thermal distributed electrons and ions. This paper is organized as follows: In Section 2, we present the basic equations governing the dynamics of our plasma in spherical coordinate. We employed the reductive perturbation method to derive Spherical Kadomtsev-Petviashvili (sKP) equation. Section 3 is devoted to the discussion of numerical analysis of soliton waves, and then we summarize the main conclusions.

## 2. Description of the model

We consider a non-planar spherical geometry with the transverse perturbation in an unmagnetized dusty plasma whose constituents are two temperature Boltzmann-distributed ions, Boltzmann-distributed electrons, and negatively charged stationary dust particles. In the equilibrium situation the charge neutrality condition becomes $Z_i n_{i0} = n_{e0} + Z_{d0} n_{d0}$, where $n_{e0}$, and $n_{d0}$ are the unperturbed electron, and dust number densities, respectively. $Z_i$, and $Z_{d0}$ are the unperturbed number of charges on the ions, and number of charges residing on the dust particles respectively. Here we consider a singly ionized plasma system for which $Z_i = 1$. An equilibrium condition for two unperturbed low ($n_{il0}$), and high ($n_{ih0}$) temperature ions density is satisfy by $n_{i0} = n_{il0} + n_{ih0}$.

The nonlinear dynamics of phase speed dust acoustic waves in spherical geometry is governed by[25, 26]

$$\frac{\partial n_d}{\partial t} + \frac{1}{r^2}\frac{\partial(r^2 n_d u_d)}{\partial r} + \frac{1}{r}\frac{\partial(n_d v_d)}{\partial \theta} + \frac{n_d v_d}{r}\cot\theta = 0 \qquad (1)$$

$$\frac{\partial u_d}{\partial t} + u_d\frac{\partial u_d}{\partial r} + \frac{v_d}{r}\frac{\partial u_d}{\partial \theta} - \frac{v_d^2}{r} = \frac{\partial \phi}{\partial r} \qquad (2)$$

$$\frac{\partial v_d}{\partial t} + u_d \frac{\partial v_d}{\partial r} + \frac{v_d}{r}\frac{\partial v_d}{\partial \theta} + \frac{u_d v_d}{r} = \frac{1}{r}\frac{\partial \phi}{\partial \theta} \tag{3}$$

$$\frac{1}{r^2}\frac{\partial}{\partial r}(r^2 \frac{\partial \phi}{\partial r}) + \frac{1}{r^2}\frac{\partial^2 \phi}{\partial \theta^2} + \frac{\cot \theta}{r^2}\frac{\partial \phi}{\partial \theta} = n_d + n_e - n_{il} - n_{ih} \tag{4}$$

Where $r$ and $\theta$ are the radial and angle coordinates and $u_d$, $v_d$ are the dust fluid velocity in $r$ and $\theta$ directions respectively. $n_d$ And $\phi$ refer to the dust number density and electrostatic wave potential respectively. The space coordinates, velocities and electrostatic potential $\phi$ are normalized by Debye length $\lambda_{De} = (T_{eff}/4\pi n_{d0} Z_{d0} e^2)^{1/2}$, the inverse of dust plasma frequency $\omega_{pd}^{-1} = (m_d / 4\pi n_{d0} Z_{d0}^2 e^2)^{1/2}$, the dust acoustic speed $C_d = (Z_{d0} T_{eff}/m_d)^{1/2}$, and $T_{eff}/e$, respectively. The effective temperature is given by

$$T_{eff} = \left( \frac{T_{il} T_{ih} T_e}{\mu_l T_{ih} T_e + \mu_h T_{il} T_e + \nu T_{il} T_{ih}} \right)$$

Where $T_{il}$ (or/and $T_{ih}$), and $T_e$ are the low (or/and high) temperature ion, and temperature electron, respectively. $\mu_l$ (or/and $\mu_h$), and $\nu$ are normalized low (or/and high) ion and electron number densities by $Z_{d0} n_{d0}$ with the intention that

$$\mu_l = \frac{\delta_1}{\delta_1 + \delta_2 - 1}, \quad \mu_h = \frac{\delta_2}{\delta_1 + \delta_2 - 1}, \quad \nu = \frac{1}{\delta_1 + \delta_2 - 1},$$

Where $\delta_1 = n_{il0}/n_{e0}$, and $\delta_2 = n_{ih0}/n_{e0}$.

We consider electrons and ions have Boltzmann distribution

$$n_e = \nu e^{\vartheta_1 s \varphi}, \quad n_{il} = \mu_l e^{-s\varphi}, \quad n_{ih} = \mu_h e^{-\vartheta s \varphi}$$

Where $\vartheta_1 = \frac{T_{il}}{T_e}, \vartheta_2 = \frac{T_{ih}}{T_e}, \vartheta = \frac{T_{il}}{T_{ih}}$ and $s = \frac{\delta_1 + \delta_2 - 1}{\delta_1 + \vartheta \delta_2 + \vartheta_1}$.

In order to investigate the propagation of dust acoustic wave in dusty plasmas, we employ the standard reductive perturbation technique to obtain the spherical Kadomtsev-Petviashvili (SKP) equation. The independent variables can be stretched as $\xi = \epsilon^{1/2}(r - v_0 t), \eta = \epsilon^{-1/2}\theta$ and $\tau = \epsilon^{3/2} t$ where $\epsilon$ is a small parameter and $v_0$ is the wave velocity. The dependent variables are expanded as

$$n_d = 1 + \epsilon n_d^{(1)} + \epsilon^2 n_d^{(2)} + \ldots \tag{5}$$

$$u_d = \epsilon u_d^{(1)} + \epsilon^2 u_d^{(2)} + \ldots \tag{6}$$

$$v_d = \epsilon^{3/2} v_d^{(1)} + \epsilon^{5/2} v_d^{(2)} + \ldots \tag{7}$$

$$\phi = \epsilon \phi^{(1)} + \epsilon^2 \phi^{(2)} + \ldots \tag{8}$$

Substituting equations (5)–(8) into the set of equations (1)–(4) and collecting terms in the different powers of $\epsilon$, to obtain then the lowest order in $\epsilon$

$$n_d^{(1)} = -\frac{1}{v_0^2}\phi^{(1)} = -\phi^{(1)} \tag{9}$$

$$u_d^{(1)} = -\frac{1}{v_0}\phi^{(1)} \tag{10}$$

$$\frac{\partial v_d^{(1)}}{\partial \xi} = -\frac{1}{v_0^2 \tau}\frac{\partial \phi^{(1)}}{\partial \eta} \tag{11}$$

From the expression of wave velocity in equation (10) we can see that the wave can propagate outward or inward. For the next higher order, we can obtain

$$\frac{\partial n_d^{(1)}}{\partial \tau} - v_0 \frac{\partial n_d^{(2)}}{\partial \xi} + \frac{\partial u_d^{(2)}}{\partial \xi} + \frac{\partial \left(n_d^{(1)} u_d^{(1)}\right)}{\partial \xi} + \frac{1}{v_0 \tau}\frac{\partial v_d^{(1)}}{\partial \eta} + \frac{1}{v_0 \tau}(2u_d^{(1)} + \frac{1}{\eta}v_d^{(1)}) = 0 \tag{12}$$

$$v_0 \frac{\partial u_d^{(2)}}{\partial \xi} - \frac{\partial u_d^{(1)}}{\partial \tau} - u_d^{(1)}\frac{\partial u_d^{(1)}}{\partial \xi} + \frac{\partial \phi^{(2)}}{\partial \xi} = 0 \tag{13}$$

$$\frac{\partial^2 \phi^{(1)}}{\partial \xi^2} = n_d^{(2)} + \phi^{(2)} + \frac{K}{2}(\phi^{(1)})^2 \tag{14}$$

Where $K = \dfrac{(\delta_1 + \delta_2 - 1)(\vartheta_1^2 - \delta_1 - \delta_2 \vartheta^2)}{(\delta_1 + \vartheta \delta_2 + \vartheta_1)^2}$

Now, using equations (9)-(11) and eliminate $n_d^{(2)}, u_d^{(2)}$, and $\phi^{(2)}$ from equations (12)-(14), one obtain the spherical Kadomtsev-Petviashvili (SKP) equation

$$\frac{\partial}{\partial \xi}[\frac{\partial \phi^{(1)}}{\partial \tau} + A\phi^{(1)}\frac{\partial \phi^{(1)}}{\partial \xi} + B\frac{\partial^3 \phi^{(1)}}{\partial \xi^3} + \frac{1}{\tau}\phi^{(1)}] + \frac{1}{2v_0 \tau^2}(\frac{\partial^2 \phi^{(1)}}{\partial \eta^2} + \frac{1}{\eta}\frac{\partial \phi^{(1)}}{\partial \eta}) = 0 \tag{15}$$

Where

$$A = -\frac{v_0}{2}(3+K) \text{ , and } B = \frac{v_0}{2}$$

If the wave propagates without the transverse perturbation, the last term in the left-hand side of equation (15) disappears and the SKP equation reduces to the ordinary spherical KdV equation. In equation (15), the two terms with variable coefficient can be cancelled if we assume

$\zeta = \xi - \frac{v_0}{2}\eta^2\tau$, $\phi^{(1)} = \phi^{(1)}(\zeta,\tau)$. Then the SKP equation (15) is reduced to the standard KdV equation

$$\frac{\partial \phi^{(1)}}{\partial \tau} + A\phi^{(1)}\frac{\partial \phi^{(1)}}{\partial \zeta} + B\frac{\partial^3 \phi^{(1)}}{\partial \zeta^3} = 0 \tag{16}$$

Following some simple algebra, the well-known solitary wave solution of the KdV equation (16) can be written as

$$\phi^{(1)} = \frac{3u_0}{A}\sec h^2[\sqrt{\frac{u_0}{4B}}(\zeta - u_0\tau)] \tag{17}$$

Where $u_0$ is a constant represent wave velocity. Thus we get an exact solitary wave solution of the SKP equation (15) as

$$\phi^{(1)} = \frac{3u_0}{A}\sec h^2[\sqrt{\frac{u_0}{4B}}\{\xi - (u_0 + \frac{v_0}{2}\eta^2)\tau\}] \tag{18}$$

### 3. Results and discussion

Figure 1 shows rarefactive solitons (peak amplitude is negative) and figure 2 shows a compressive soliton corresponding to the positive peak amplitude. These Figures show the solution of spherical KdV equation for several values of $\tau$ ranging from $\tau = 3$ to $\tau = 9$. As indicated in these figures, the amplitude decreases with increasing values of $\tau$ and it shows that if we increase $\tau$ to large values, the nonplanar geometries would approach the planar geometry.

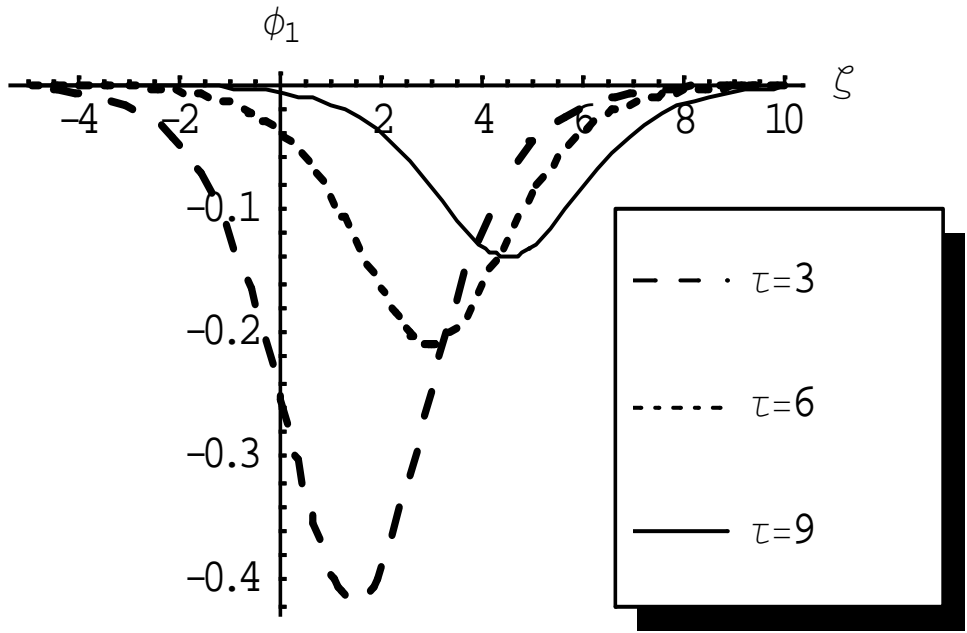

Figure 1: Electrostatic potential $\phi^{(1)}$ vs. $\zeta$ with $\tau = 3$, $\tau = 6$ and $\tau = 9$

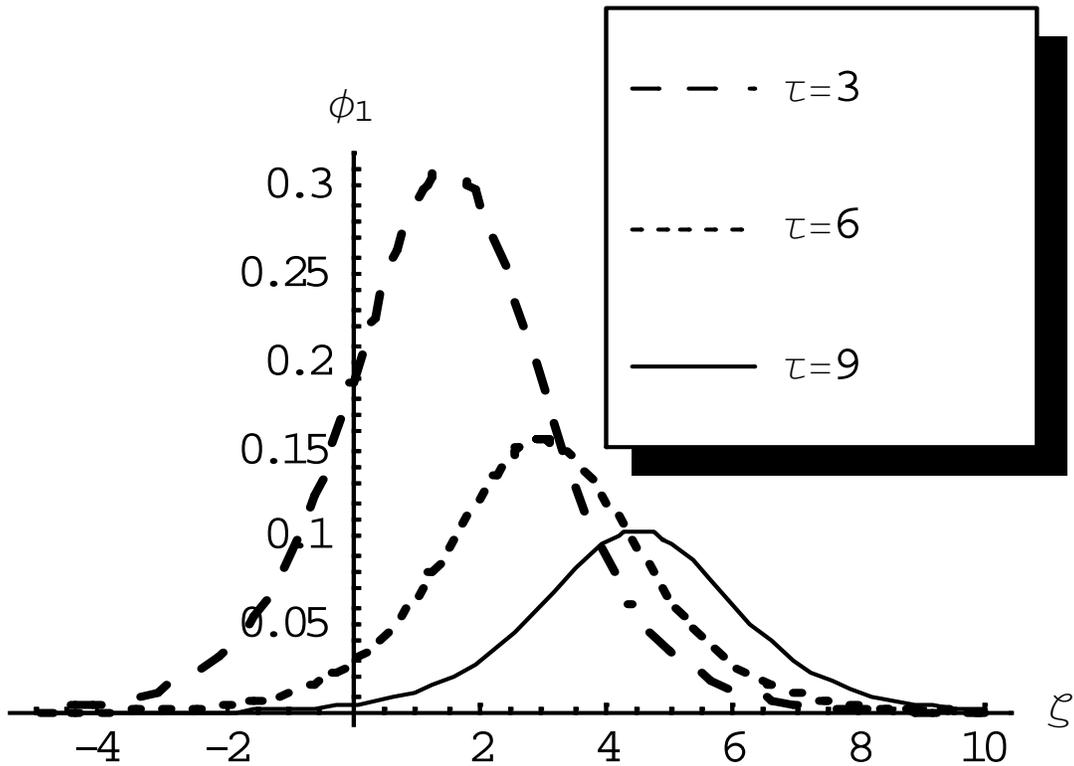

Figure 2: Electrostatic potential $\phi^{(1)}$ vs. $\zeta$ with $\tau = 3$, $\tau = 6$ and $\tau = 9$

In summary, we have derived a spherical KP equation for a spherical dust acoustic wave in unmagnetized dusty plasmas by employing the perturbation technique. We have found the exact solution for the SKP equation for DA solitary waves. It is shown that a solitary wave can exist in a bounded spherical geometry under the transverse perturbation. It is clear that the amplitude and wave velocity of our solitary wave described by SKP equation (15) are exclusively determined by the parameters of the system and only depending on the initial conditions. It should be noted that the coefficients of dispersive and nonlinear terms $A$ and $B$ get modified due to the presence of quantum effects. These terms contribute to the formation of amplitude and width of spherical dust acoustic solitary waves. For a large value of $\tau$, equation (15) reduces to ordinary Kdv equation and the spherical solitary waves can be similar to planar geometry solution. Furthermore, it must be noted that both types of solitons are admissible. As $u_0 > 0$ it is clear from (17) or (18) that if $A > 0$, there exists solitary waves with positive potential or compressive solitary waves and if $A < 0$, there exists solitary waves with negative potential or rarefactive solitary waves.

It is worth to mention that these results are in three dimensional geometry which is a realistic situation in laboratory or space plasmas. We, therefore, propose to perform a laboratory experiment which can study such special new features of dust acoustic solitary waves.